\documentclass{article}
\title{WordSig: QR streams enabling platform-independent self-identification that's impossible to deepfake
\\ }
\usepackage{natbib}
\usepackage[utf8]{inputenc} 
\usepackage[T1]{fontenc}    

\usepackage{hyperref}       
\PassOptionsToPackage{hyphens}{url}\usepackage{hyperref}
\usepackage{hyperref}
\hypersetup{pdftex,colorlinks=true,allcolors=blue}
\usepackage{hypcap}
\usepackage{url}            

\usepackage{booktabs}       
\usepackage{amsfonts}       
\usepackage{nicefrac}       
\usepackage{microtype}      
\usepackage{amsmath}
\usepackage{amsthm}
\usepackage{amssymb}
\usepackage{float}
\usepackage{graphicx}
\usepackage{xcolor}
\usepackage{amsmath}
\usepackage{amsthm}
\usepackage{amssymb}
\usepackage{wrapfig}
\usepackage[capitalise]{cleveref}

\makeatletter
\newcommand{\vast}{\bBigg@{3}}
\newcommand{\Vast}{\bBigg@{4}}
\makeatother

\makeatletter
\newcommand{\bidef}[3]{
  \protected@write\@auxout{}{%
    \global\string\@namedef{#1:#2}{#3}%
  }%
}

\newcommand{\biref}[3]{
  \@ifundefined{#1:#2}{??}{\@nameuse{#1:#2}}%
}
\makeatother

\usepackage{todonotes}
\usepackage{soul}
\usepackage{marginnote} %

\newif\ifcomments 

\newif\iffootnote
\let\Footnote\footnote
\renewcommand\footnote[1]{\begingroup\footnotetrue\Footnote{#1}\endgroup}

\newcommand\setupcomments[4]{
\expandafter\def\csname#1says\endcsname##1{\ifcomments{\color{#3}[#2: ##1]}\fi}
\expandafter\def\csname#1predicts\endcsname##1{\ifcomments{\color{#3}[#2 predicts: ##1 \#}\fi}
\expandafter\def\csname#1color\endcsname##1{\ifcomments{\color{#3}##1}\fi}
\expandafter\def\csname#1adds\endcsname##1{\ifcomments{\color{#3}##1}\fi}
\expandafter\def\csname#1rems\endcsname##1{\ifcomments{\color{#3}\st{##1}}\fi}
  \expandafter\def\csname#1box\endcsname##1{\ifcomments\vspace{1ex}\todo[inline,#4]{#2 says: ##1}{}\fi}

  \expandafter\def\csname#1margin\endcsname##1{\ifcomments\marginnote{\todo[inline,#4]{#2 says: ##1}{}}\fi}
  \expandafter\def\csname#1section\endcsname##1{\ifcomments\todo[#4]{#2 says: ##1}{}\fi}
  \expandafter\def\csname#1todo\endcsname##1{\ifcomments\marginnote{\todo[inline]{#2 says: TODO: ##1}{}}\fi}

  \expandafter\def\csname#1demo\endcsname{\ifcomments
\noindent
$\backslash$#1says $\to$ \csname#1says\endcsname{example inline comment from #2.}\\
$\backslash$#1adds $\to$ \csname#1adds\endcsname{example addition by #2.}\\
$\backslash$#1rems $\to$ \csname#1rems\endcsname{example removal by #2.}\\
$\backslash$#1box $\to$ 
    \csname#1box\endcsname{example box comment from #2}
\fi}

}

\commentstrue
\usepackage[linesnumbered,ruled]{algorithm2e}
\SetAlFnt{\small}
\SetAlCapFnt{\normalsize}
\SetAlCapNameFnt{\normalsize}
\SetArgSty{textrm} 
\SetKw{KwContext}{Context:}
\SetKwBlock{Def}{def}{enddef}
\newcommand{\QRcode}{\mathrm{QRcode}}
\newcommand{\ReadQRcode}{\mathrm{ReadQRcode}}

\newcommand{\Priv}{\mathrm{Priv}}
\newcommand{\Cert}{\mathrm{Cert}}
\newcommand{\Pub}{\mathrm{Pub}}
\newcommand{\Sig}{\mathrm{Sig}}
\newcommand{\Trusted}{\mathrm{Trusted}}

\newcommand{\display}{\textrm{display}\xspace}
\SetKwFor{Def}{def}{:}{enddef}
\usepackage{framed}
\usepackage{quoting}
\colorlet{shadecolor}{blue!10}
\newenvironment{shadedquotation}
 {\begin{shaded*}
  \quoting[leftmargin=0pt, vskip=0pt]
 }
 {\endquoting
 \end{shaded*}
}

\usepackage{comment}
\usepackage{float}

\usepackage{scrextend} 
\usepackage{environ} 

\NewEnviron{story}[3]{
\vspace{1ex}
\label{story:#1}\bidef{story}{#1}{#2}
\begin{shadedquotation}
\textbf{#3}
\BODY
\end{shadedquotation}
\vspace{1ex}
}
\newcommand{\storyref}[1]{\hyperref[story:#1]{\biref{story}{#1}}}

\commentstrue
\setupcomments{ac}{AC}{purple!100!black}{color=blue!30}
\setupcomments{sr}{SR}{green!70!black}{color=green!10}

\usepackage{array}
\newcolumntype{L}[1]{>{\raggedright\let\newline\\\arraybackslash\hspace{0pt}}p{#1}}
\newcolumntype{C}[1]{>{\centering\let\newline\\\arraybackslash\hspace{0pt}}p{#1}}
\newcolumntype{R}[1]{>{\raggedleft\let\newline\\\arraybackslash\hspace{0pt}}p{#1}}

\author{
  Andrew Critch\thanks{Center for Human-Compatible Artificial Intelligence, Department of Electrical Engineering and Computer Sciences, UC Berkeley} \\
  \texttt{critch@eecs.berkeley.edu} \\
}

\begin{document}

\maketitle

\begin{abstract}
Deepfakes can degrade the fabric of society by limiting our ability to trust video content from leaders, authorities, and even friends.  Cryptographically secure digital signatures may be used by video streaming platforms to endorse content, but these signatures are applied by the content distributor rather than the participants in the video.  We introduce WordSig, a simple protocol allowing video participants to digitally sign the words they speak using a stream of QR codes, and allowing viewers to verify the consistency of signatures across videos.  This allows establishing a trusted connection between the viewer and the participant that is not mediated by the content distributor.  Given the widespread adoption of QR codes for distributing hyperlinks and vaccination records, and the increasing prevalence of celebrity deepfakes, 2022 or later may be a good time for public figures to begin using and promoting QR-based self-authentication tools.
\end{abstract}


\section{Introduction} 

\citet{westerlund2019emergence} provides a fine introduction to the concept and problem of deepfakes:

\begin{quote}
Recent technological advancements have made it easy to create what are now called “deepfakes”, hyper-realistic videos using face swaps that leave little trace of manipulation \citep{chawla2019deepfakes}. Deepfakes are the product of artificial intelligence (AI) applications that merge, combine, replace, and superimpose images and video clips to create fake videos that appear authentic \citep{maras2019determining}. Deepfake technology can generate, for example, a humorous, pornographic, or political video of a person saying anything, without the consent of the person whose image and voice is involved \citep{day2019future,fletcher2018deepfakes}. The game-changing factor of deepfakes is the scope, scale, and sophistication of the technology involved, as almost anyone with a computer can fabricate fake videos that are practically indistinguishable from authentic media \citep{fletcher2018deepfakes}. While early examples of deepfakes focused on political leaders, actresses, comedians, and entertainers having their faces weaved into porn videos \citep{hasan2019combating}, deepfakes in the future will likely be more and more used for revenge porn, bullying, fake video evidence in courts, political sabotage, terrorist propaganda, blackmail, market manipulation, and fake news \citep{maras2019determining}.

While spreading false information is easy, correcting the record and combating deepfakes are harder \citep{roets2017fake}.

\end{quote}

This article presents an approach to the deepfake problem that enables users to self-identify in videos, in a way that would be cryptographically difficult or impossible to fake.  The idea is that a user appearing in a video can use their own mobile device to digitally sign their words as they speak, and display the signatures are a series of QR codes that are cryptographically impossible to generate without the user's private key, but can be verified for authenticity and consistency with their words using the user's public key.  In particular, the user can self-identify in this way without any support or cryptographic tooling from the person(s) recording the video, or the platform(s) distributing the video.

Thus, the WordSig approach to deepfakes is to enable the creation of verifiably-non-fake content, specifically, streams of words digitally signed with QR codes.  This proactive approach is complementary to approaches that \emph{detect} deepfake videos after the videos have been produced \citep{guera2018deepfake, li2018exposing}.  These after-the-fact detection approaches run the risk of obsolescence by the development of ever more effective techniques for fabricating content.  By contrast, the proactive digital signature technique present here is based on time-tested cryptographic techniques that are well known to be computationally intractable to fake.  

Still, proactive techniques like WordSig are only a partial solution to the problem of deepfakes.  For instance, when fake videos are produced to accuse people of immoral or embarrassing acts, the videos will not typically depict those persons voluntarily displaying WordSig IDs (or any form of ID) to identify themselves to the camera.  Such videos will need to be debunked as fake by whatever the latest after-the-fact detection techniques happen to exist at the time.  Nonetheless, one can hope that important public figures could eventually adopt the habit of always self-identifying with WordSig (or similar) when they make an important public announcement, so that important statements like ``We are (not) at war'' that \emph{are not} accompanied by a verifiable WordSig stream could be distrusted by default, like a legal agreement without a signature.

\section{Deployment}

In real-world deployment, a WordSig QR stream would appear as depicted in Figures \ref{figure:tryscanning} and \ref{figure:wordsigid} below.

\begin{figure}[!ht]
  \caption{Example to scan}\label{figure:tryscanning}
  \vspace{1ex}
  \includegraphics[width=\textwidth]{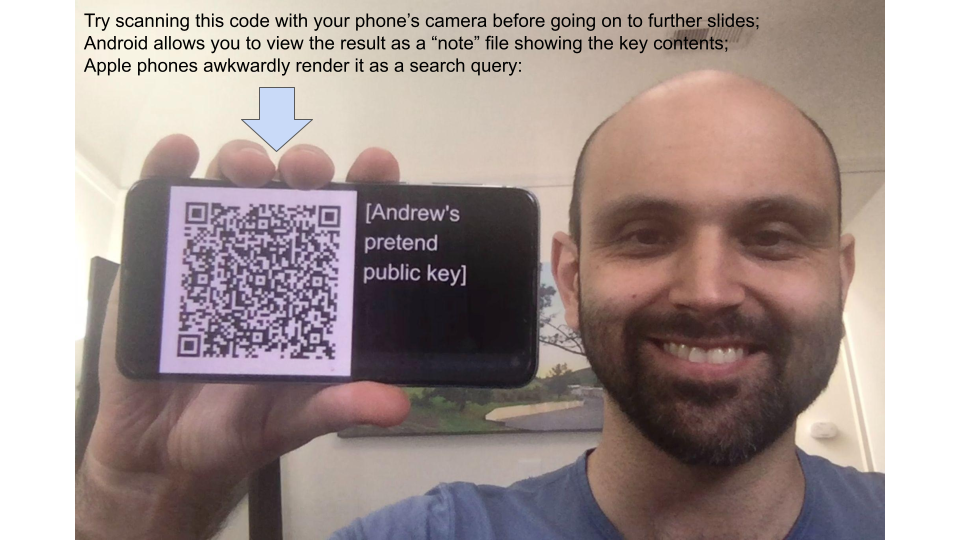}
\end{figure}

\begin{figure}[!ht]
  \caption{Another example to scan}\label{figure:wordsigid}
  \vspace{1ex}
  \includegraphics[width=\textwidth]{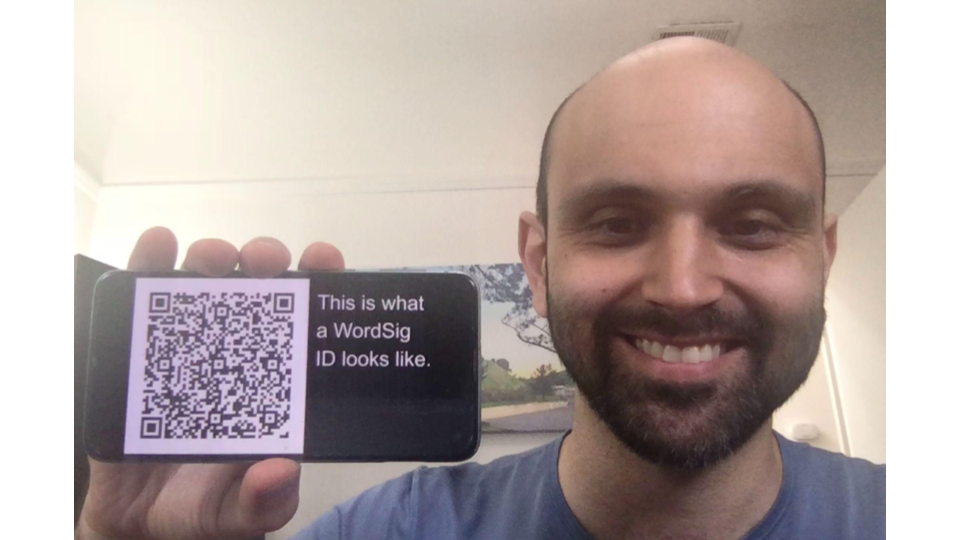}
\end{figure}

\section{Algorithm}

The algorithm for writing the WordSig code stream enmeshes spoken content with signed content, so that editors are unable to copy and paste words out of context of the preceding words:

\begin{figure}[!ht]
  \caption{Structure of the QR stream}\label{figure:flow}
  \vspace{1ex}
  \includegraphics[width=\linewidth]{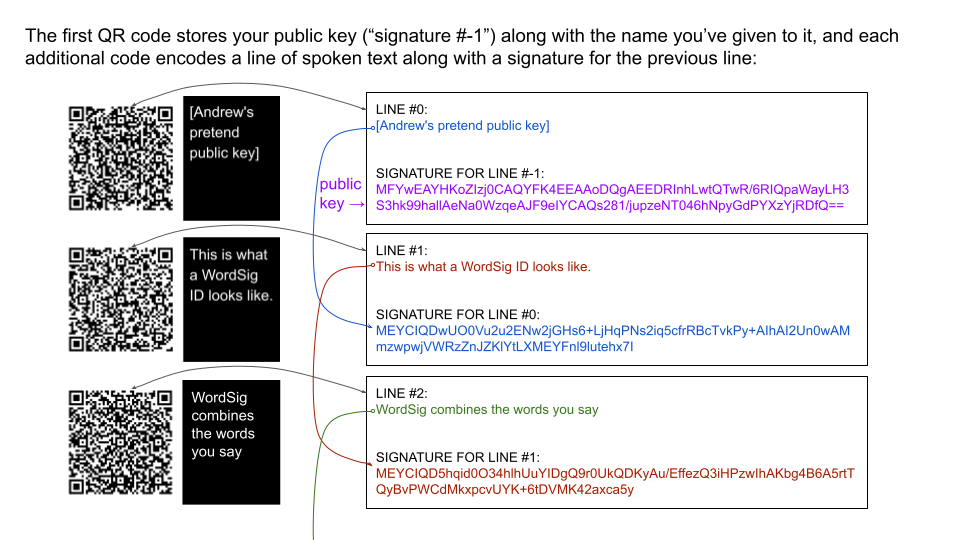}
\end{figure}

\begin{algorithm}[!ht]
 \caption{Signing \& streaming}
 \KwContext{A user JaneDoe123 speaking into her mobile device and displaying its screen to a video camera.}\\
 \KwIn{A stream of words spoken by Jane, grouped into 2-second-long segments, $W = (w_1, w_2, \ldots$)}
 \KwIn{Jane's Secp256k1 private key $\Priv$ stored securely on her mobile device.}
 \KwIn{Jane's public key certificate $\Cert$, containing her public key $\Pub$, signed with Jane's private key $\Priv$ and perhaps other users or authorities.} 
 \KwOut{A stream of QR codes $Q = (q_0, q_1, \ldots)$ certifying Jane's words using her private key $\Priv$.}
 $q_0 = \QRcode(\Cert)$\\
 $w_0 = ``[\textrm{JaneDoe123's public key certificate}]"$\\
 $\display(q_0,w_0)$\\
 \For{i = 1,2,3,\ldots}{
  $\Sig_{i-1} = \mathrm{ECDSA\_Sign}(w_{i-1},\Priv)$\\
  $q_i = \QRcode(\mathrm{concatenate}(w_{i},\textrm{``::''},\Sig_{i-1}))$\\
  $\display(q_i,w_i)$
 }
\end{algorithm}

\begin{algorithm}[!ht]\footnotesize
 \caption{Verifying the stream}
 \KwContext{A user Kevin Doe watching a video in which Jane speaks and displays her QR code signature stream from Algorithm 1.}\\
 \KwIn{A stream of QR codes $(q_0,q_1,\ldots)$ and word groups $(w_0,w_1,\ldots)$ from Jane}
 \KwIn{A history of previously trusted certificates $\Trusted$, stored on Kevin's device.}
 \KwIn{(optional) A (possibly empty) database Revoked containing certificates labeled and signed as ``Revoked'' by their own private keys.}

 \KwOut{Assessment of message validity.}
  $\Cert = \ReadQRcode(q_0)$\\
  $\mathrm{CertName} = \mathrm{ExtractName}(\Cert)$;   $\mathrm{TextName} = \mathrm{ExtractName}(w_0)$\\
  \tcp{These should both equal JaneDoe123:}
  \uIf{TextName $\neq$ CertName}{
    \Return ``Fake: text name does not match certificate name''\\
    }
  \uIf{not ECDSA\_Verify\_Certificate(Cert)}{
    \Return ``Fake: certificate signature does not match certificate name.''\\
  }
  \uIf{$\Cert \in \mathrm{Revoked}$}{
    \Return ``Possibly fake: Certificate was revoked using its own private key on'' + ExtractDateRevoked(Cert,Revoked)\\
  }
  \Def{QueryForTrust()}{
      \display ``Do you trust that this content is from'' + TextName + ``?''\\
    response = queryViewer()\\
     \eIf{response == True}{
     append(Trusted, Cert)
     }{
     \Return ``Possibly fake: you do not trust the certificate source."
     }
 }
  \uIf{$\Cert \not\in \Trusted$}{
    QueryForTrust() 
  } 
  \uIf{$\mathrm{CertName} \in \Trusted$ and LatestTrustedCertificateFor(CertName) $\neq \Cert$}{
    \display ``Certificate does not match latest trusted certificate for'' + CertName + ``;''\\
    \display ``Possibly fake signature stream.''\\
    QueryForTrust()
  }

  \For{i>0}{
    $w'_i = \mathrm{ExtractText}(\ReadQRcode(q_i))$\\
    $\Sig_{i-1} = \mathrm{ExtractSignature}(\ReadQRcode(q_i))$\\
    \uIf{not $w_i = w'_i$}{
    \Return ``Fake: QR code text content does not match displayed text content.''\\
    }
    \uIf{not ECDSA\_Verify\_Signature($Cert,w_{i-1},\Sig_{i-1}$)}{
    \Return ``Fake: Signature'' + (i-1) + ``does not match words and certificate.''\\
    }
    \display ``Signatures verified thus far...''
  }
  \Return ``Signature stream verified.''
\end{algorithm}
 
\section{Adoption and promotion by public figures}
Announcements like ``We are not at war with {[country]}'' or ``We are at war with [country]'' could be particularly important to receive and authenticate or debunk in a timely manner.  How can we make these messages easier to distribute in a broadly verifiable way?

Leaders in government and other important public service institutions could use WordSig to authenticate important announcements, such as in press conferences where many news agencies are recording a video of the same scene.  By displaying the digital signature in the scene being recorded, there would be no need to establish cryptographic standards amongst the various large and small news agencies recording the announcement.    However, it's notoriously difficult to convince the government or political leaders to use any particular kind of technology, perhaps due to an accumulated resistance to private institutions engaging in lobbying.  So, perhaps public service applications will not be the easiest source of early usage for this technology, despite it being readily applicable and useful.

Celebrities, on the other hand, could be a great source of early adopters.  Celebrities have a natural individual incentive to protect their personal brand, and are generally more free to promote new technologies than members of government.  So, early real-world implementations of WordSig should probably be aimed at use cases that appeal to celebrities, while demonstrating capabilities that will be useful for public service applications in government and NGOs.  Since this year (2022) has marked considerable advancements in AI-generated web content (both text and images), now might be a good time to begin encouraging celebrities to self-authenticate using WordSig or similar digital signature techniques.

\section{Limitations and remedies}
While the WordSig algorithm is helpful for empowering users to self-authenticate in a platform-independent way, it remains an incomplete remedy to the problem of deepfakes:

\begin{enumerate}
    \item Users who assume WordSig authenticates the entire video, rather than just the words being spoken and signed, will not be protected.  For instance, if Police Chief Alice Smith points at a man named Bob Taylor standing next to her in a video and says ``This man is not a fugitive'', the video can be edited to replace Bob Taylor by another man —-- perhaps one who is a fugitive --- but Alice's signature on her spoken words would remain valid.  Alice needs to say ``Bob Taylor is not a fugitive''.  This is why we call the technique ``WordSig'': as a reminder that only the spoken words are being signed and authenticated.
    \item Users who choose to trust new WordSig certificates for no reason will not be protected from deception.  This could be addressed by ensuring WordSig app(s) will remind users not to assume authenticity, and by the development of a network of certificate authorities following protocols similar to HTTPS certificate authorities.
    \item Users who automatically trust videos that display a WordSig ID, without scanning the video with a WordSig reader to verify it, will not be protected.  Thus, it helps if they have a friend who would remind them to actually scan the video, or if mobile devices would automatically suggest scanning WordSig content the way they suggest scanning QR codes in photos.  Celebrities who choose to use WordSigs to generate authentic content could also remind users to scan the content.
    \item Users who install fake apps that pretend to be WordSig scanners will not be protected.  So, app stores and mobile device providers will need to prevent the proliferation of fake scanners.
\end{enumerate} 

\section{Summary and Discussion}

QR code streams that embed spoken words and digital signatures are proposed to empower individuals to represent themselves in videos in a manner that is impossible to mimic with deepfake techniques.  The name ``WordSig'' is proposed for this technique, to remind users that the words (and only the words) are what is being signed.  A mock-up of a WordSig signature stream is presented, and a case is made for celebrities endorsing and promoting WordSig as a way of securing their own brands and identities, which in turn could lead to political and government leaders making similar use of the technology to authenticate official communications with the public.  In order for WordSig to be effective, WordSig verifier apps will need to ensure users understand what the app can and cannot be used to verify, and mobile device providers will need to prevent the spread of fake apps pretending to be WordSig verifiers.

\cleardoublepage
\bibliography{main}

\begin{thebibliography}{}

\bibitem[\protect\citeauthoryear{Chawla}{Chawla}{2019}]{chawla2019deepfakes}
Chawla, R. (2019).
\newblock Deepfakes: How a pervert shook the world.
\newblock {\em International Journal of Advance Research and
  Development\/}~{\em 4\/}(6), 4--8.

\bibitem[\protect\citeauthoryear{Day}{Day}{2019}]{day2019future}
Day, C. (2019).
\newblock The future of misinformation.
\newblock {\em Comput. Sci. Eng.\/}~{\em 21\/}(1), 108.

\bibitem[\protect\citeauthoryear{Fletcher}{Fletcher}{2018}]{fletcher2018deepfakes}
Fletcher, J. (2018).
\newblock Deepfakes, artificial intelligence, and some kind of dystopia: The
  new faces of online post-fact performance.
\newblock {\em Theatre Journal\/}~{\em 70\/}(4), 455--471. Johns Hopkins
  University Press.

\bibitem[\protect\citeauthoryear{G{\"u}era and Delp}{G{\"u}era and
  Delp}{2018}]{guera2018deepfake}
G{\"u}era, D. and E.~J. Delp (2018).
\newblock Deepfake video detection using recurrent neural networks.
\newblock In {\em 2018 15th IEEE international conference on advanced video and
  signal based surveillance (AVSS)}, pp.\  1--6. IEEE.

\bibitem[\protect\citeauthoryear{Hasan and Salah}{Hasan and
  Salah}{2019}]{hasan2019combating}
Hasan, H.~R. and K.~Salah (2019).
\newblock Combating deepfake videos using blockchain and smart contracts.
\newblock {\em Ieee Access\/}~{\em 7}, 41596--41606. IEEE.

\bibitem[\protect\citeauthoryear{Li and Lyu}{Li and Lyu}{2018}]{li2018exposing}
Li, Y. and S.~Lyu (2018).
\newblock Exposing deepfake videos by detecting face warping artifacts.
\newblock {\em arXiv preprint arXiv:1811.00656\/}.

\bibitem[\protect\citeauthoryear{Maras and Alexandrou}{Maras and
  Alexandrou}{2019}]{maras2019determining}
Maras, M.-H. and A.~Alexandrou (2019).
\newblock Determining authenticity of video evidence in the age of artificial
  intelligence and in the wake of deepfake videos.
\newblock {\em The International Journal of Evidence \& Proof\/}~{\em 23\/}(3),
  255--262. SAGE Publications Sage UK: London, England.

\bibitem[\protect\citeauthoryear{Roets et~al.}{Roets
  et~al.}{2017}]{roets2017fake}
Roets, A. et~al. (2017).
\newblock ‘fake news’: Incorrect, but hard to correct. the role of
  cognitive ability on the impact of false information on social impressions.
\newblock {\em Intelligence\/}~{\em 65}, 107--110. Elsevier.

\bibitem[\protect\citeauthoryear{Westerlund}{Westerlund}{2019}]{westerlund2019emergence}
Westerlund, M. (2019).
\newblock The emergence of deepfake technology: A review.
\newblock {\em Technology Innovation Management Review\/}~{\em 9\/}(11).

\end{thebibliography}
\end{document}